# Unified Network-Based Representation of BIM Models for Embedding Semantic, Spatial, and Topological Data

Jin Han[1], Xin-Zheng Lu[1], and Jia-Rui Lin[1,*]

[1]Department of Civil Engineering, Tsinghua University, China
han-j23@mails.tsinghua.edu.cn, luxz@tsinghua.edu.cn, lin611@tsinghua.edu.cn

**Abstract –**
Building Information Modeling (BIM) has revolutionized the construction industry by providing a comprehensive digital representation of building structures throughout their lifecycle. However, existing research lacks effective methods for capturing the complex spatial and topological relationships between components in BIM models, which are essential for understanding design patterns and enhancing decision-making. This study proposes a unified network-based representation method that integrates the "semantic-spatial-topological" multi-dimensional design features of BIM models. By extending the IFC (Industry Foundation Classes) standard, we introduce local spatial relationships and topological connections between components to enrich the network structure. This representation method enables a more detailed understanding of component interactions, dependencies, and implicit design patterns, effectively capturing the semantic, topological, and spatial relationships in BIM, and holds significant potential for the representation and learning of design patterns.

## 1 Introduction

Building Information Modeling (BIM) is a digital methodology employed in construction to design, manage, and improve detailed 3D models of buildings across their entire lifecycle [1]. By consolidating detailed information about a structure's physical and functional properties into one unified model, BIM has greatly enhanced the efficiency and effectiveness of engineering applications [2]. Consequently, BIM has been widely studied in various fields, including visualization [3], human-machine collaborative design [4] [5], and life cycle assessment [6] [7], and its adoption is steadily increasing. However, the inherent complexity of BIM models, which contain vast amounts of semantic, spatial, and topological information, makes it challenging to leverage their full potential for automated analysis and decision-making [8]. Traditional methods of working with BIM models often struggle to capture the nuanced relationships between components, such as spatial proximities, functional dependencies, and topological connections, hindering advanced tasks like design validation, conflict detection, and knowledge extraction. Therefore, how to extract and represent relevant design features and patterns from BIM data to enhance the design, construction, and maintenance of buildings has become a key focus of scholarly research.

To address the demand for a standardized and interoperable framework in construction, the Industry Foundation Classes (IFC) were introduced as an open data format to facilitate the exchange of BIM information across various tools [9] [10]. However, IFC's complex structure and irrelevant data for many tasks have led researchers to represent BIM models as vectors or networks [11]. Some studies have focused on extracting key component attributes, converting them into tensors for machine learning applications, such as clash detection and classification [12]-[14]. While these approaches address component attributes, they overlook important spatial and topological relationships, which are crucial for understanding component interactions. To capture these relationships, other researchers have modeled BIM components as nodes and edges in a network, representing attributes as node or edge features [15][17]. While these studies address some limitations of the first method, the spatial and topological relationships they capture are often simplistic and insufficient. Furthermore, the graph construction process is typically tailored to specific tasks, lacking flexibility and general applicability.

Therefore, this study proposes a unified network-based representation method for the BIM model's "semantic-spatial-topological" multi-dimensional design features. Building upon IFC and previous research, it further incorporates local spatial relationships and topological relationships between components to enrich the local information within the network. This approach not only facilitates the integration of different data types but also enables the identification of implicit design patterns, thereby enhancing the understanding of component interactions and dependencies.

## 2 Background

To address the growing demand for a standardized and interoperable framework in the construction industry, IFC were introduced [9]. IFC was developed as an open, neutral data format to ensure the exchange of information across various BIM tools without data loss or misinterpretation [10]. However, due to the non-intuitive nature of the data in IFC and the presence of irrelevant information for most tasks, many researchers have attempted to further represent BIM models as vectors, or networks based on IFC and BIM software.

The first type primarily involves extracting key attributes of components and converting them into independent tensors for training and prediction. Lin & Huang (2019) extracted data from clash detection reports, encoded it into tensors, and used machine learning to detect unrelated conflicts [12]. Wang & Leite (2013) selected three types of attributes—geometric properties (e.g., "Volume"), usage and materials (e.g., "Material Flexibility"), and clash information (e.g., "Clashing Volume"), which were converted into tensors and used to train machine learning algorithms for predicting clash responses [13]. Liu et al. (2024) used component images and basic attributes like thickness and concrete grade to train a multi-model deep learning system for classification [14]. Although the above studies considered key component attributes, their use of tensors overlooked important spatial and topological relationships, such as relative positions, connections, and collaborative functions (e.g., "force transfer mechanisms"). These relationships are not only key to understanding component interactions and functional coordination but also crucial in capturing design patterns. For instance, the spatial relationship between beams and columns reflects how they transfer forces together. Without this information, models cannot fully capture the implicit interactions between components.

Therefore, scholars have attempted to represent BIM models as networks, where components and their relationships are depicted as nodes and edges, and attributes are represented as node or edge features. For example, Wang et al. (2022) treat point-based instances (e.g., "Pipe Fitting") and curve-based instances (e.g., "Pipe Curve") as nodes and edges in a network, using graph matching techniques to detect connection errors between components [15]. Hu et al. (2020; 2023) represent spatial and topological relationships, such as clashes and connections, as edges for collision detection and optimization [5] [16]. In contrast to the fine-grained representation, Li et al. (2024) treat rooms as nodes, with edges representing their hierarchical relationships, enabling automatic modular building design [17]. While these methods effectively capture certain spatial and topological properties of BIM models, they show considerable variation in network construction depending on the task, lacking a standardized approach to BIM model representation. Moreover, the absence of spatial relationships between components leads to incomplete information in the network.

## 3 Unified Network-based Representation Method for BIM

While the IFC data format can describe the geometry, attributes, and relationships of building components in detail, its storage format involves complex hierarchical structures and a vast amount of semantic information. However, deep learning networks, which are commonly used for data analysis, require numerical, flattened formats (such as images, matrices, or vectors) as input. Therefore, in order to leverage the data within existing BIM models for analysis, learning, and reuse, it is necessary to extract and preprocess the data from IFC to simplify its structure and extract key features.

In this study, we propose a unified network-based representation method for BIM models that retains critical information while structuring the data to accommodate various BIM models and research questions. Specifically, using IFC files exported from BIM software, this study extracts and calculates the properties of various components and their spatial and topological relationships, as illustrated in Table 1. Subsequently, all components are then represented as different types of nodes based on their classification in the IFC, with the extracted semantic properties represented as node features. Meanwhile, the spatial and topological relationships between components are represented as edges connecting the corresponding nodes. This network-based representation method lays the foundation for subsequent learning and transfer of design patterns within BIM models.

### 3.1 Semantic Data Representation of Components

Specifically, in representing the semantic properties of components, this study selects only the attributes related to each component itself from the IFC files exported by the BIM model, including basic information, geometric shapes, material properties, and other relevant attributes. The basic information includes attributes such as component ID, name, and FamilyName, which are used to uniquely identify the component. Geometric shapes are described by the attribute "Representation," including BoundingBox, Brep, SectionedSpine, and others. Material properties are stored in entities such as "IfcMaterial" and "IfcMaterialLayerSetUsage," describing the composition and characteristics of the component. Other relevant attributes are defined through "IfcPropertySet", including durability, thermal conductivity, compressive strength, and more.

### 3.2 Topological Data Representation of Components

For topological properties, we selected the host and connection relationships contained in the IFC. In IFC, entities such as doors (IfcDoor) and windows (IfcWindow) are typically associated with wall entities (IfcWallStandardCase) through an opening entity (IfcOpeningElement). However, in this study, the opening entity is ignored, and the host relationship is directly associated with the wall and window or door entities. Besides, connection relationships are obtained through "IFCRELCONNECTSPATHELEMENTS," which describe the connections between components.

In addition, this study introduces a touch floor relationship to describe the vertical connection between components and floors, providing a more intuitive representation of the functional layout and component distribution across different building levels. For example, this includes the connection between walls and floors, or the installation positions of equipment within a floor relative to the floor slab. This topological relationship is extracted by calculating the positional relationship between the bounding boxes of components and floors. The three types of topological relationships extracted and calculated are shown in Figure 1.

Table 1 Unified attributes extraction and calculation method for semantic, topological, and spatial properties of components

| Properties | Attributes | Description |
|---|---|---|
| Semantic | Basic information | Including attributes such as component ID, name, and FamilyName, which are used to identify the component uniquely. |
| | Geometric shapes | Defined by the attribute "Representation" from IFC, including BoundingBox, Brep, and others, this is used to represent the geometric shape of components. |
| | Material properties | Defined by attributes such as "IfcMaterial" and "IfcMaterialLayerSetUsage" in IFC, describing the composition and material of the component. |
| | Other relevant attributes | Defined by "IfcPropertySet", including durability, thermal conductivity, compressive strength, and more. |
| Spatial | Location | Coordinates of component positioning points or positioning line endpoints. It is defined by "ObjectPlacement," including the local coordinate system and reference coordinate system. |
| | Spatial relationships | The positional relationship between two components in space, including five categories: 1) different surface, 2) interface non-parallel, 3) interface parallel, 4) point-to-line, and 5) point-to-point. |
| | Angle | The intersection angle between two components' positioning lines; if one component uses point positioning, the angle is 0. |
| | Shortest distance vector coordinates | Endpoint coordinates of the shortest distance vector between the positioning points or lines of two components. |
| | Shortest distance | Actual spatial separation between two components. A negative value indicates an overlap or collision between the components, while a positive value denotes a gap. |
| | Angle with the horizontal plane | The angle between the plane or line formed by the two components and the horizontal plane. For spatial relationships in categories 2-5, the angle with the horizontal plane represents the angle between the plane or line formed by their positioning lines or points and the horizontal plane. In category 1, the angle is 0. |
| Topological | Host relationship | Represents the inclusion relationship, where one component is fully contained within another, such as doors/windows within walls. |
| | Connection relationship | Defined by "IFCRELCONNECTSPATHELEMENTS," which describe the connections between components. |
| | Touch floor | Describe the vertical connection between components and floors. |

relationship

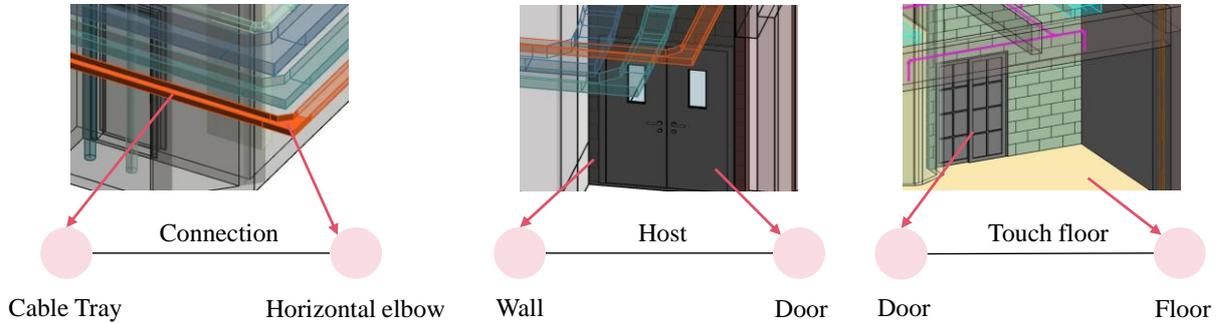

Figure 1. Three types of topological relationships

### 3.3 Spatial Data Representation of Components

This study first extracts the attributes representing the location of components from the IFC files, which are described by "ObjectPlacement," including the local coordinate system and reference coordinate system.

However, simple coordinate information alone is insufficient to intuitively reflect the spatial relationships between components. To address this limitation, this study further incorporates local spatial relationships of nearby components to provide additional geometric information and contextual understanding. Specifically, a local spatial relationship identifies all components within a specified distance (0.5m in this study) from a given component and calculates their relative positional relationships, including: 1) spatial relationship, 2) angle, 3) shortest distance vector coordinates, 4) shortest distance, and 5) angle with the horizontal plane, as illustrated in Figure 2.

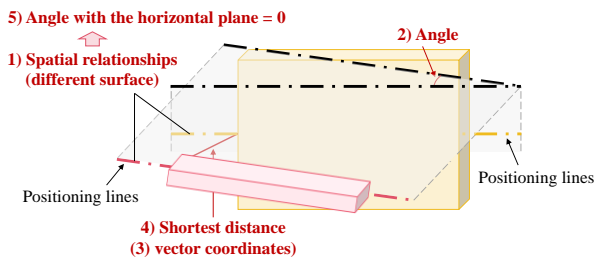

Figure 2. Example of local spatial relationship

The "spatial relationship" refers to the positional relationship between two components in space, which is determined by their positioning points or lines, and it can be classified into five distinct types: 1) different surfaces, 2) non-parallel interfaces, 3) parallel interfaces, 4) point-to-line, and 5) point-to-point. The first three types apply when both components are positioned using lines, the fourth type applies when one component is positioned by points and the other by lines, and the fifth applies when both components are positioned by points. The "angle" refers to the intersection angle between the positioning lines of the two components. If one component uses point positioning, the angle is considered 0°. The "shortest distance vector coordinates" correspond to the endpoints of the shortest distance between the positioning points or lines of the two components. The "distance" value indicates the spatial separation between the components. A negative distance suggests overlap or collision, while a positive value indicates a gap, showing that the components are separated by a specific distance. The "angle with the horizontal plane" refers to the angle between the plane or line formed by the two components and the horizontal plane. For spatial relationships falling under categories 2 to 5, the "angle with the horizontal plane" is calculated as the angle between the plane or line formed by the positioning points/lines and the horizontal plane. For category 1, the angle is 0°. Based on geometric principles, it is clear that these five calculated spatial relationships can comprehensively describe the relative positional relationship between two components.

### 3.4 Network-based Representations of BIM Models

This section further converts the components, their attributes, and relationships extracted from Sections 2.1, 2.2, and 2.3 into network nodes, node features, edges, and edge features, to achieve the network representation of the BIM model. The construction process is outlined in Table 2.

Specifically, all components in the BIM model are treated as nodes in the network. Based on the classification of component categories in IFC, we treat these nodes into different types, such as walls, beams, pipes, and so on. At the same time, all semantic properties of the components extracted in Section 2.1 are converted into the corresponding node features. The topological relationships extracted in Section 2.2 are represented as edges connecting the corresponding component nodes. These edges are further classified into three categories based on the three different types of topological relationships. Additionally, the spatial relationships between a component and all components within a local 0.5m range, as described in Section 2.3, are represented as the fourth type of edge in the network. The calculated relative positional relationships between two components are used as the features of this type of edge.

It should be noted that, when two component nodes are connected by a topological relationship, they are inherently adjacent, with a relative distance of zero. This proximity implies an inherent spatial relationship between them, which could lead to redundancy if both spatial and topological relationships are represented as separate edges. To prevent this redundancy, we adopted a strategy where the spatial relationship is disregarded when a topological relationship is present, ensuring that the two component nodes are connected by only one edge. This methodology effectively reduces redundancy while maintaining the integrity of both the spatial and topological relationships within the network.

Table 2. Process of constructing the BIM-specific network

| Algorithm |
|---|
| Input: Components $\mathcal{C}$ from BIM model $\mathcal{M}$ |
| Output: Network $\mathcal{N}\{\mathcal{V}, \mathcal{E}\}$; node features $\{\mathbf{n}_v, \forall v \in \mathcal{V}\}$; edge features $\{\mathbf{n}_e, \forall e \in \mathcal{E}\}$ |
| 1:   For each component type $\mathcal{T}$ in $\mathcal{C}$ do |
| 2:     For each component $i$ in $\mathcal{T}$ do |
| 3:       $v_i^{\mathcal{T}}$ = create_node($i$) |
| 4:       $\mathbf{n}_{v_i^{\mathcal{T}}}$ = extract_semantic_ properties($i$) |
| 5:     end for |
| 6:   end for |
| 7:   For each node $v_i^{\mathcal{T}}$ in $\mathcal{V}$ do |
| 8:     Nodes $v_{all\_j}^{\mathcal{T}}$ = get_host_relationship($i$) |
| 9:     For $v_j^{\mathcal{T}}$ in $v_{all\_j}^{\mathcal{T}}$: |
| 10:       $(v_i^{\mathcal{T}}, v_j^{\mathcal{T}})_h$ = create_host_edge($v_i^{\mathcal{T}}, v_j^{\mathcal{T}}$) |
| 11:     end for |
| 12:     Nodes $v_{all\_k}^{\mathcal{T}}$ = get_connection_relationship($i$) |
| 13:     For $v_k^{\mathcal{T}}$ in $v_{all\_k}^{\mathcal{T}}$: |
| 14:       $(v_i^{\mathcal{T}}, v_k^{\mathcal{T}})_c$ = create_connection_edge($v_i^{\mathcal{T}}, v_k^{\mathcal{T}}$) |
| 15:     end for |
| 16:     Nodes $v_{all\_l}^{\mathcal{T}}$ = get_touch_floor_relationship($i$) |
| 17:     For $v_l^{\mathcal{T}}$ in $v_{all\_l}^{\mathcal{T}}$: |
| 18:       $(v_i^{\mathcal{T}}, v_l^{\mathcal{T}})_T$ = create_touch_floor_edge($v_i^{\mathcal{T}}, v_l^{\mathcal{T}}$) |
| 19:     end for |
| 20:     For each node $v_m^{\mathcal{T}}$ in $\mathcal{V}$ and $m \neq i$: |
| 21:       if (distance between $m$ and $i$ < spatial relationship threshold) and (not $(v_i^{\mathcal{T}}, v_m^{\mathcal{T}})_h \& (v_i^{\mathcal{T}}, v_m^{\mathcal{T}})_c \& (v_i^{\mathcal{T}}, v_m^{\mathcal{T}})_T$): |
| 22:         $(v_i^{\mathcal{T}}, v_m^{\mathcal{T}})_S$ = create_spatial_edge($v_i^{\mathcal{T}}, v_m^{\mathcal{T}}$) |
| 23:         $\mathbf{n}_{(v_i^{\mathcal{T}}, v_m^{\mathcal{T}})_S}$ = calculate_relative_position($i$, $m$) |
| 24:     end for |
| 25: end for |

## 4 Case Study

Our network-based representation method achieves seamless integration with mainstream BIM software by leveraging the standardized IFC format as the universal data interface. Since IFC serves as the open BIM interoperability standard supported by all major platforms (Revit, ArchiCAD, etc.), our method is theoretically applicable to nearly all BIM software. Users simply need to export the model as an IFC file and use the proposed program for conversion. This section uses a residential floor as a case study to present how the method proposed in this research can be applied to convert a BIM model into a network. The resulting overall network and the zoomed-in local network are shown in Figure 3. Notably, node positions in the figure are determined by the coordinates of their corresponding components' central points.

As shown in Figure 3, representing a BIM model as a network effectively simplifies the complex building information and clearly expresses the multi-dimensional features of components and their relationships, such as semantic, spatial, and topological relationships. Additionally, the introduction of the touch floor relationship and the spatial relationships within a 0.5m local range further enhances the network by adding more local relational information. Overall, this method effectively compresses redundant geometric data while retaining as much of the hierarchical component attributes and spatial-temporal dependencies as possible, resulting in a structured yet information-rich network representation.

With the rapid development of machine learning, particularly in the context of nonlinear and fuzzy learning, there is an increasing need for structured tabular data for training. Our network-based BIM model addresses this requirement by providing a clear and structured format that can be used for training machine learning models, including graph neural networks (GNNs). Given that our model incorporates numerous spatial and topological features, it offers a powerful foundation for learning and reusing design features from historical models. For instance, GNNs can be employed to mask and reconstruct node features in an unsupervised manner, enabling the model to deeply capture implicit design patterns within the BIM data. This allows the model to understand complex relationships and interactions between components that are not explicitly labeled but are critical for tasks like design optimization, automated checks, and conflict detection. Additionally, our method is directly applicable to downstream tasks. For example, in automated checks and conflict detection, we can convert these tasks into edge relationship prediction tasks, where the goal is to predict the relationships between two nodes (components) in the network. Using GNNs, we can predict whether any potential conflicts or design issues exist based on the spatial and topological relationships encoded in the network. This approach not only enables accurate detection but also facilitates the optimization and validation of designs in a more automated and efficient manner.

In summary, the network-based representation we propose supports machine learning algorithms, especially GNNs, by providing a structured format that captures both explicit and implicit design features. This facilitates tasks such as conflict detection, design optimization, and more, by making the data more accessible and interpretable for downstream analysis.

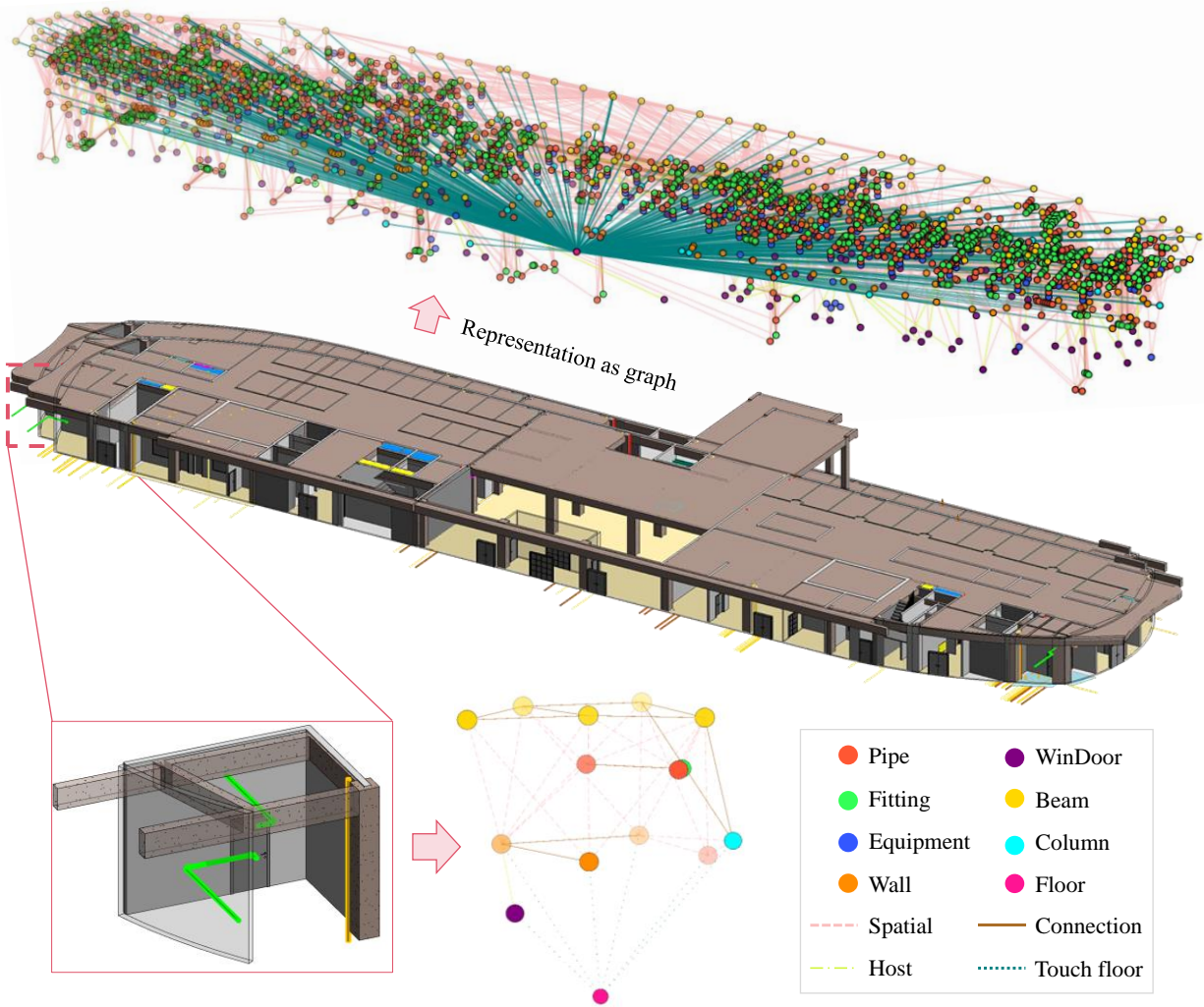

Figure 3. Example of local spatial relationship

## 5 Conclusion

This study proposes a unified network-based representation method for BIM models, effectively capturing the semantic, spatial, and topological features of building components. By transforming IFC data into a structured network format, our approach not only simplifies the representation but also enhances its utility for machine learning and other advanced analytical tasks. The proposed method enables more efficient design feature extraction, conflict detection, and optimization, offering a promising framework for leveraging BIM data in various downstream applications.

Beyond the scope of the current research, future improvements can be pursued through the following approaches:

(a) Additional component features, such as time or design phase, can be integrated into the node and edge attributes to account for the impacts of construction processes and design stages, thus broadening the scope of downstream transfer learning tasks. For instance, in spatiotemporal clash detection, properties like construction priority and design phase can be incorporated. Similarly, for operation and maintenance applications, attributes such as access space and frequency of access can be included.

(b) By applying advanced machine learning techniques such as GNNs to the building data converted into a network, we can learn and capture implicit design patterns within the BIM data, enabling the transfer and reuse of historical data.

## Acknowledgments

The authors are grateful for the financial support received from the National Key R&D Program of China (No. 2023YFC3804600) and the National Natural Science Foundation of China (No. 52378306).